\documentclass[
 reprint,
superscriptaddress,blueshift
showpacs,preprintnumbers,
amsmath,amssymb,
aps,
prb,
]{revtex4-1}

\usepackage{graphicx}%
\usepackage{dcolumn}%
\usepackage{bm}
\usepackage{color} 
\usepackage{comment}
\includecomment{comment}
\begin{document}

\title{Direct measurement of polariton-polariton interaction strength in the Thomas-Fermi regime of exciton-polariton condensation.}

\author{E. Estrecho}
\affiliation{Nonlinear Physics Centre, Research School of Physics and Engineering, The Australian National University, Canberra, ACT 2601, Australia}
\affiliation{ARC Centre of Excellence in Future Low-Energy Electronics Technologies}
\author{T. Gao}
\affiliation{Institute of Molecular Plus, Tianjin University, 300072 Tianjin, China}
\author{N. Bobrovska}
\affiliation{Institute of Physics, Polish Academy of Sciences, A. Lotinik\'ow 32$/$46, 02-668 Warsaw, Poland}
\author{D. Comber-Todd}
\affiliation{Nonlinear Physics Centre, Research School of Physics and Engineering, The Australian National University, Canberra, ACT 2601, Australia}
\author{M. D. Fraser}
\affiliation{JST, PRESTO, 4-1-8 Honcho, Kawaguchi, Saitama 332-0012, Japan}
\affiliation{Quantum Functional System Research Group, RIKEN Center for Emergent Matter Science, 2-1 Hirosawa, Wako-shi, Saitama 351-0198, Japan}
\author{M. Steger}
\affiliation{National Renewable Energy Lab, Golden, CO 80401, USA}
\author{K. West}
\affiliation{Princeton Institute for the Science and Technology of Materials (PRISM), Princeton University, Princeton, New Jersey 08544, USA}
\author{L. N. Pfeiffer}
\affiliation{Department of Electrical Engineering, Princeton University, Princeton, New Jersey 08544, USA}
\author{J. Levinsen}
\affiliation{School of Physics and Astronomy, Monash University, Melbourne, VIC 3800, Australia}
\affiliation{ARC Centre of Excellence in Future Low-Energy Electronics Technologies}
\author{M. M. Parish}
\affiliation{School of Physics and Astronomy, Monash University, Melbourne, VIC 3800, Australia}
\affiliation{ARC Centre of Excellence in Future Low-Energy Electronics Technologies}
\author{T. C. H. Liew}
\affiliation{Division of Physics and Applied Physics, Nanyang Technological University, Singapore}
\author{M. Matuszewski}
\affiliation{Institute of Physics, Polish Academy of Sciences, A. Lotinik\'ow 32$/$46, 02-668 Warsaw, Poland}
\author{D. W. Snoke}
\affiliation{Department of Physics and Astronomy, University of Pittsburgh, PA 15260, USA}
\author{A. G. Truscott}
\affiliation{Laser Physics Centre, Research School of Physics and Engineering, The Australian National University, Canberra, ACT 2601, Australia}
\author{E. A. Ostrovskaya}
\affiliation{Nonlinear Physics Centre, Research School of Physics and Engineering, The Australian National University, Canberra, ACT 2601, Australia}
\affiliation{ARC Centre of Excellence in Future Low-Energy Electronics Technologies}

\begin{abstract}
Bosonic condensates of exciton polaritons (light-matter quasiparticles in a semiconductor) provide a solid-state platform for studies of non-equilibrium quantum systems with a spontaneous macroscopic coherence. These driven, dissipative condensates typically coexist and interact with an incoherent reservoir, which undermines measurements of key parameters of the condensate. Here, we overcome this limitation by creating a high-density exciton-polariton condensate in an optically-induced ÒboxÓ trap. In this so-called Thomas-Fermi regime, the condensate is fully separated from the reservoir and its behaviour is dominated by interparticle interactions. We use this regime to directly measure the polariton-polariton interaction strength, and reduce the existing uncertainty in its value from four orders of magnitude to within three times the theoretical prediction. The Thomas-Fermi regime has previously been demonstrated only in ultracold atomic gases in thermal equilibrium. In a non-equilibrium exciton-polariton system, this regime offers a novel opportunity to study interaction-driven effects unmasked by an incoherent reservoir.  
\end{abstract}

\maketitle
\section{Introduction}

Exciton polaritons (polaritons herein) are hybrid light-matter bosons formed by strongly interacting photons and excitons in semiconductor microcavities \cite{Microcavity_book}. Sufficiently strong off-resonant optical pumping can drive a spontaneous transition of polaritons to Bose-Einstein condensation \cite{Deng_02,BEC06,BEC07,Deng_10,CiutiREV13,YamamotoREV14}. The details of this transition are strongly influenced by the reservoir of high-energy excitonic particles created and maintained by the optical pump. The role of the reservoir is two-fold: first and foremost, it provides a source of particles that form the condensate via stimulated scattering processes. Secondly, it creates a local potential barrier for polaritons due to the energy shift induced by strong, repulsive interactions  between condensing polaritons and thermal reservoir particles. This feature has been successfully employed to create a vast variety of reconfigurable pump-induced potentials for polaritons \cite{Bloch10,Deveaud11,Tosi12,Tosi_vortices,Cristofolini13,Askitopoulos13,Dall14,Baumberg14,Gao15,Liew_ring,Bayer15,Snoke15,Review17,Sun18}. Tailored engineering of the optical potential enables Bose-Einstein condensation of polaritons in a trap \cite{Cristofolini13,Askitopoulos13,Liew_ring,Sun18} and precise manipulation of polariton flows \cite{Bloch10,Bayer15}. Several fundamental properties of polariton systems have been explored using this technique, ranging from the formation of persistent currents \cite{Tosi_vortices,Dall14,Baumberg14} to geometric phases associated with Hermitian \cite{Eliezer16} and non-Hermitian spectral degeneracies (exceptional points) \cite{Gao15}.
Many of these studies rely on the fact that optically trapped polariton condensates are driven and typically exhibit multimode behaviour (fragmentation \cite{fragmentation06,fragmentation07}), i.e., they occupy several excited single-particle energy eigenstates in the effective potential \cite{multimode2,gain_guiding_exp,multimode3,Bloch11,Tosi12}. These features are common for trapped, highly non-equilibrium bosonic condensates, and have also been observed in condensates of photons \cite{photonBEC} and magnons \cite{magnonBEC}.

Here, we present a reliable method for creating single-mode, high-density polariton condensates in the ground state of an optically induced trap. The method utilises strong depletion of the reservoir (spatial hole burning) \cite{Eliezer17} in the regime of pulsed excitation of long-lifetime polaritons \cite{SnokePRX13,Snoke}, which `burns out' a circular `box' potential well for polaritons and fully removes the reservoir from the spatial region occupied by the condensate. Using this method, we create a high-density, trapped polariton condensate in the Thomas-Fermi regime, which has never been observed in this intrinsically non-equilibrium system. Previous studies of confined polariton condensates in a defect-induced trap \cite{multimode3} have revealed the deformation of the ground state wavefunction and approach towards the Thomas-Fermi regime in the multimode region of polariton condensation. By contrast, here we observe single-mode condensates, which are spatially separated from the reservoir and exhibit a uniform density and chemical potential. In analogy with uniform clouds of ultracold atoms \cite{BECbox}, these condensates lend themselves to the Thomas-Fermi approximation, which states that the chemical potential of a high-density condensate in the ground state of a confining trap is uniquely determined by the condensate mean-field energy \cite{Stringari96,Pethick96}. The interaction-dominated Thomas-Fermi regime in a box potential achieved in our experiments allows us to perform a direct measurement of the polariton-polariton interaction strength. We compare our results with theoretical predictions \cite{ciuti98,tassone,Rochat,Glazov}, and put them in the context of existing measurements above \cite{Amo09,Bloch11,Forchel11,Rodriguez16,Walker17} and below \cite{Nelson17} the condensation threshold.

\section{Condensation in the Thomas-Fermi regime}

Spontaneous Bose-Einstein condensation of polaritons in a two-dimensional circular trap formed by an annular continuous wave ({\em cw}) optical pump has been explored in detail \cite{Cristofolini13,Askitopoulos13,Liew_ring,Sun18}, and is useful for creating condensates that are spatially separated from the reservoir. In our experiments, the trap is formed by shaping the pulsed laser beam into a ring which is then imaged onto the sample (see Appendix A). Consequently, a strong repulsive ring-shaped barrier is induced by the interaction between the pump-injected excitonic reservoir and condensing polaritons. The resulting circular trap has a diameter ranging from approximately $10$ to $50$ $\mu$m, and has a constant barrier width around $4~\mu$m. The condensate forms inside the trap and later decays on the timescale of the polariton lifetime ($\sim200$ ps). The pulsed excitation (see Appendix A) ensures that the excitonic reservoir (with lifetime of at least an order of magnitude longer than that of polaritons) is not continuously replenished, and hence the spatial depletion of the reservoir can become significant \cite{Eliezer17}. 

\begin{figure*}[ht]
\centering
\includegraphics[width=0.8\textwidth]{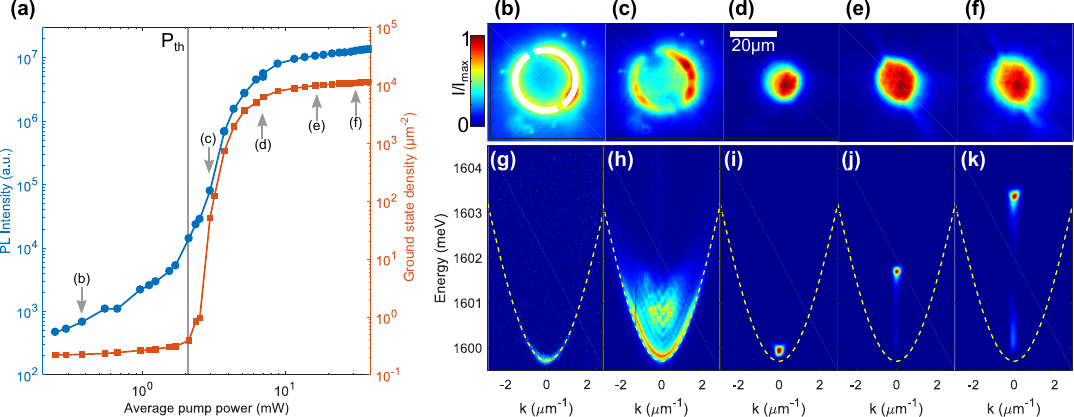}
\caption{\textbf{High-density single-mode condensation.}
(a) Pump power series of total photoluminescence intensity (blue) and ground state density of polaritons (orange). $P_{th}$ is the pump threshold for condensation.
(b-f) Real space distribution and the corresponding (g-k) dispersion (spatially filtered emission from inside the trap) of polaritons at increasing pump power corresponding to labeled points in (a).
The white overlay in (b) depicts the excitation profile, while the dashed lines in (g-k) correspond to the lower polariton dispersion.
The photon-exciton detuning is $\Delta=+2$~meV and the ring diameter is $D=25$~$\mu$m. Typical data for negative (photonic) detunings are presented in Section A of the Supplemental Material\cite{SM}.
}
\label{dispersion}
\end{figure*}

Typical signatures of the transition to condensation in the ground state of the trapping potential are presented in Fig.~\ref{dispersion}. The condensation threshold is signalled by a sharp growth in the emission intensity from the polaritons in the lowest energy state at the pump power threshold $P_{\rm th}$, as demonstrated in Fig.~\ref{dispersion}(a). At very low pump powers, below the condensation threshold, the thermal polaritons display a typical parabolic dispersion [Fig. \ref{dispersion}(g)]. Above the threshold, condensation occurs into a fragmented state characterised by a macroscopic occupation of several energy eigenstates in the effective trapping potential, as shown in  Fig.~\ref{dispersion}(h). The large occupation of high-energy, high-momenta states leads to an appreciable overlap with the pump region, as seen in the real-space image, Fig.~\ref{dispersion}(c). With growing pump power, the stimulated scattering and phonon-assisted relaxation processes drive condensation towards lower-order energy states of the trap until a single-mode occupation of the ground state is achieved, which is accompanied by a 4--5 orders of magnitude increase in density. This high-density condensate has a narrow momentum space distribution [Fig. \ref{dispersion}(i-k)] and is well confined inside the trap [see Fig.~\ref{dispersion}(d-f)]. The condensate energy blueshifts with a further increase in pump power, as shown in Fig.~\ref{dispersion}(j,k). The dispersion images in Fig.~\ref{dispersion} capture the entire lifetime of each condensate realisation after the pulsed excitation. As the condensate decays, its density decreases, which is accompanied by a decrease in the blueshift. This process manifests itself as a low-energy tail in the time-integrated images [e.g. Fig. \ref{dispersion}(j,k)].

The ability of polaritons to condense into the ground state of the system $E_{\rm min}({\bf k}=0)$ strongly depends on the detuning between the cavity photon energy and the exciton resonance, $\Delta=E_c-E_{X}$, which determines the proportion of photon and exciton in the polariton, as well as the strength of the polariton-polariton interaction \cite{Forchel11,CiutiREV13} and the relevant decay rates \cite{Vina04,Eliezer17}. For polaritons with a large photonic fraction (large negative detuning), occupation of low-energy states, including the ground state, may never be achieved as seen in spatially resolved energy measurements [Fig.~\ref{detuning}(a,b)] and the dispersion shown in the Supplemental Material \cite{SM}, Section A. 
By contrast, polaritons with a high excitonic fraction (positive detuning) readily undergo the transition into the
 ground state for moderate above threshold pump powers, as seen in Fig. \ref{detuning}(c,d) and (e,f). However, the larger effective mass leads to short-range propagation of excitonic polaritons. As a result, higher pump powers are needed to accumulate polaritons in the centre of the trap. This creates a large density in the excitonic reservoir leading to tightening of the trap as demonstrated in Fig. \ref{detuning}(e,f) for large positive detuning.

The striking difference between the spatially filamented condensate density corresponding to a fragmented condensate [Fig. \ref{dispersion}(h) and Fig. \ref{detuning}(a,b)] and a smooth, weakly deformed density of a condensate in the ground state [Fig. \ref{dispersion}(d-f),(i-k) and Fig. \ref{detuning}(c,d), (e,f)] is revealed by single-shot imaging in real space, as seen in Section B of the Supplemental Material\cite{SM}. Similar to the case of photon-like condensates formed under Gaussian excitation conditions \cite{Eliezer17}, the fragmented condensates display large density fluctuations and shot-to shot variations that persist with increasing pump power well above threshold.  In addition, fragmented condensates leak outside the trap due to the non-negligible population of weakly confined high-energy states. By contrast, single-mode condensates are well confined, and display spatially smooth density profiles with minimal shot-to-shot fluctuations.

\begin{figure*}[ht]
\centering
\includegraphics[width=0.8\textwidth]{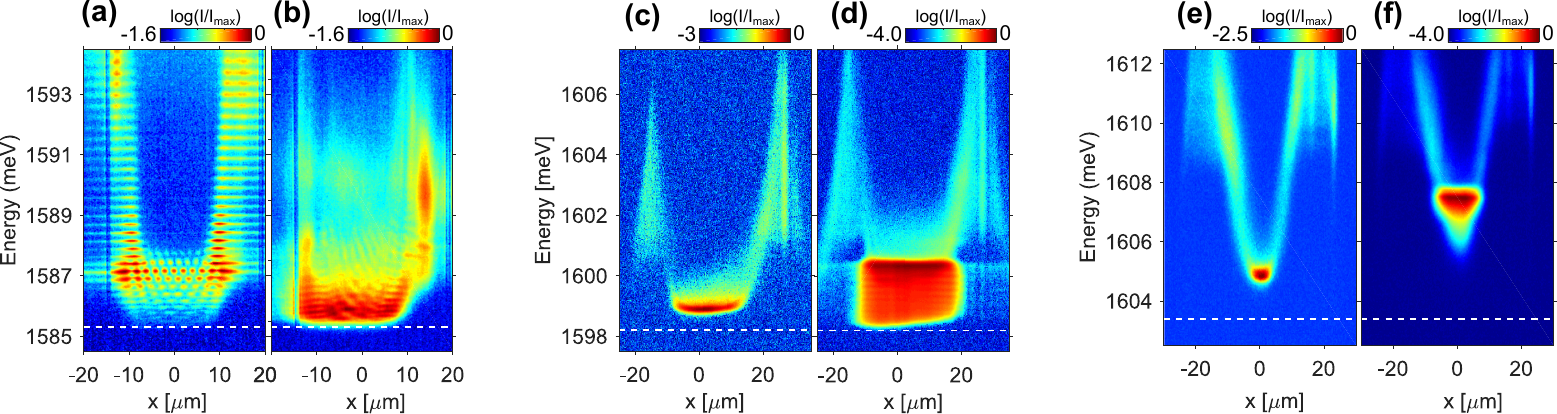}
\caption{\textbf{Dependence of condensation on detuning.}
(a,b) Spatially resolved energy measurements of multi-mode condensation of highly photon-like polaritons ($\Delta = -18$~meV) showing occupation of multiple excited states both at (a) just above threshold and (b) at the highest available pump power.
(c,d) Ground state condensation of near-resonance polaritons ($\Delta = 0$~meV) showing a small initial blueshift (c) just above threshold and an efficient reservoir depletion at (d) higher pump power as evidenced by the bottom of the energy tail, which is also the bottom of the trapping potential, being lower than the initial blueshift.
(e,f) Ground state condensation of highly exciton-like polaritons ($\Delta = +20$~meV) showing a (e) large initial blueshift due to tight trapping at threshold and (f) inefficient reservoir depletion evidenced by the higher bottom of the energy tail.
The ring pump in (a,b) ($D=20$ $\mu$m) and (e,f) ($D=30$ $\mu$m) was created using metal masks, while the $D=40$~$\mu$m trap in (c,d) was created using an axicon lens.
Dashed lines are the energy minima of polaritons in the low-density limit, $E^0_{LP}$. The intensity is plotted in logarithmic scale to accentuate relaxation of the energy to the bottom of the trap.}
\label{detuning}
\end{figure*}

\begin{figure*}[ht]
\centering
\includegraphics[width=0.8\textwidth]{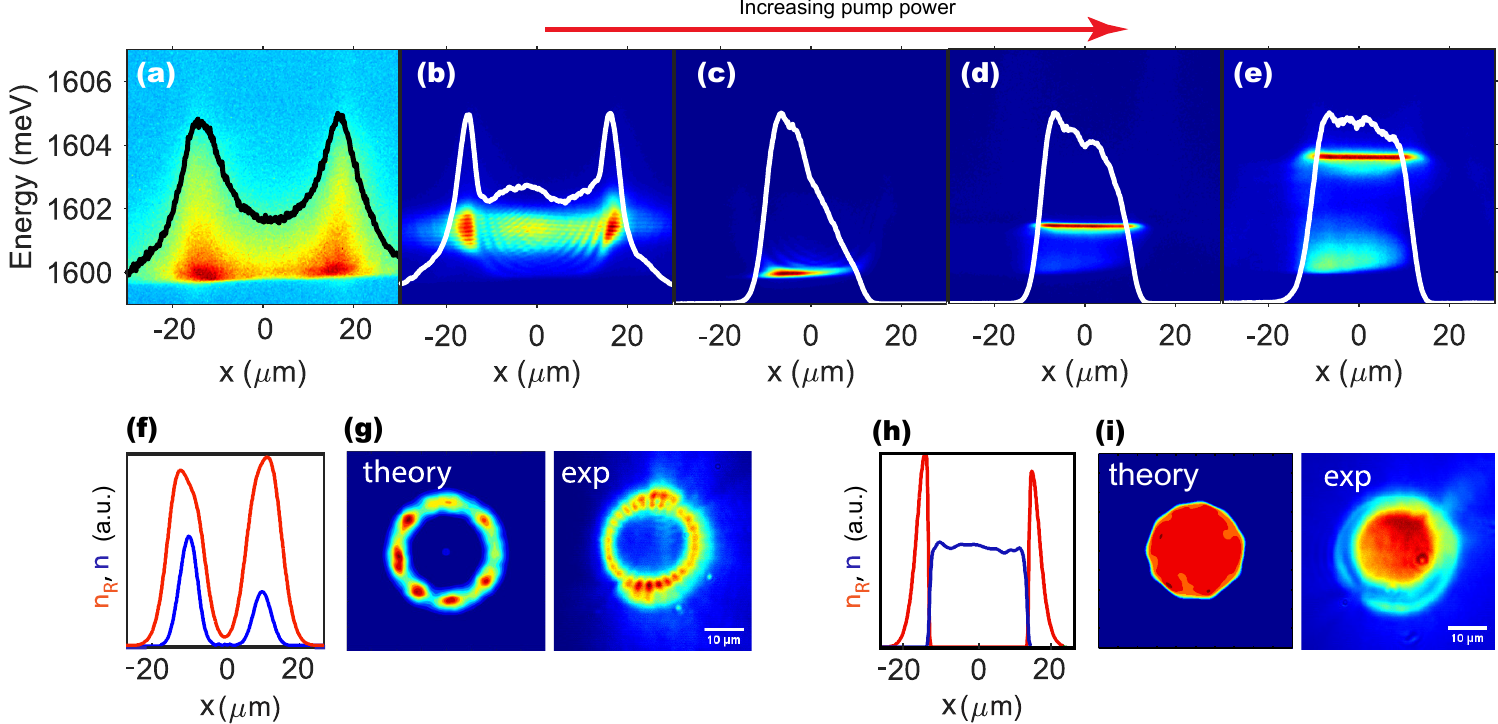}
\caption{\textbf{Transition to a flat-top condensate profile in a circular box potential.}
(a-e) Spatial profile (solid lines) of the polariton density in the ground state extracted from single-shot real-space images for $\Delta = +2$~meV and $D=30$ $\mu$m (a) below and (b-e) above condensation threshold, at the respective pump powers: $0.17P_{\rm th}$, $1.2P_{\rm th}$, $2.1P_{\rm th}$, $7.7P_{\rm th}$, $20P_{\rm th}$.
Panels (f) and (h) show numerically calculated spatial densities of the reservoir $n_{\rm R}$ (red) and polaritons $n$ (blue) corresponding to the excitation conditions in (b) and (e), respectively. The condensate density in (f) is magnified by a factor of 5. The corresponding real space density distributions found numerically and measured experimentally are shown in panels (g,i). See Appendix B for details of the numerical modeling.} 
\label{profiles}
\end{figure*}

\begin{figure*}[t]
\centering
\includegraphics[width=0.8\textwidth]{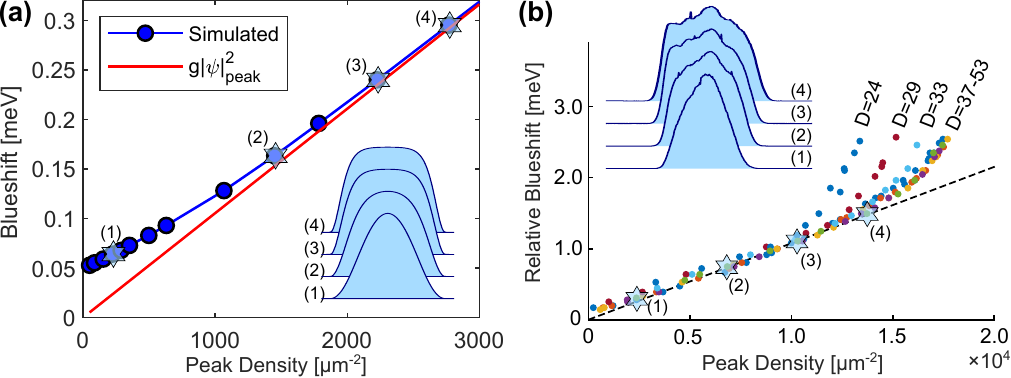}
\caption{ \textbf{Measurement of the polariton-polariton energy blueshift.}
(a) Simulated blueshift (blue points with the line as a guide to the eye) as a function of peak density for a polariton condensate in a fixed $D=20~\mu$m circular box potential for $g = 0.105~\mu$eV$\cdot \mu$m$^2$. Red solid line corresponds to the linear Thomas-Fermi regime due to polariton--polariton interactions. Inset: Normalised spatial profiles corresponding to the starred data points.
(b) Experimental blueshift-density curves for condensates in optically-induced circular traps of various diameters $D$ (measured in $\mu$m) at the detuning of $\Delta = +0.7$~meV showing a linear Thomas-Fermi regime followed by a further increase in blueshift at high densities.
Dashed line is the linear fit to the data points that correspond to a flat-top (or steep-edge) condensate profile, which is consistently achieved for trapping diameters $D \geq 29~\mu$m.
The overall behaviour is independent of the trap sizes for $D \geq 37~\mu$m. Inset: Normalised spatial profiles of the condensate density corresponding to the marked data points in the linear regime for $D=37~\mu$m.}
\label{blueshift}
\end{figure*}

\begin{figure*}[t]
\centering
\includegraphics[width=0.8\textwidth]{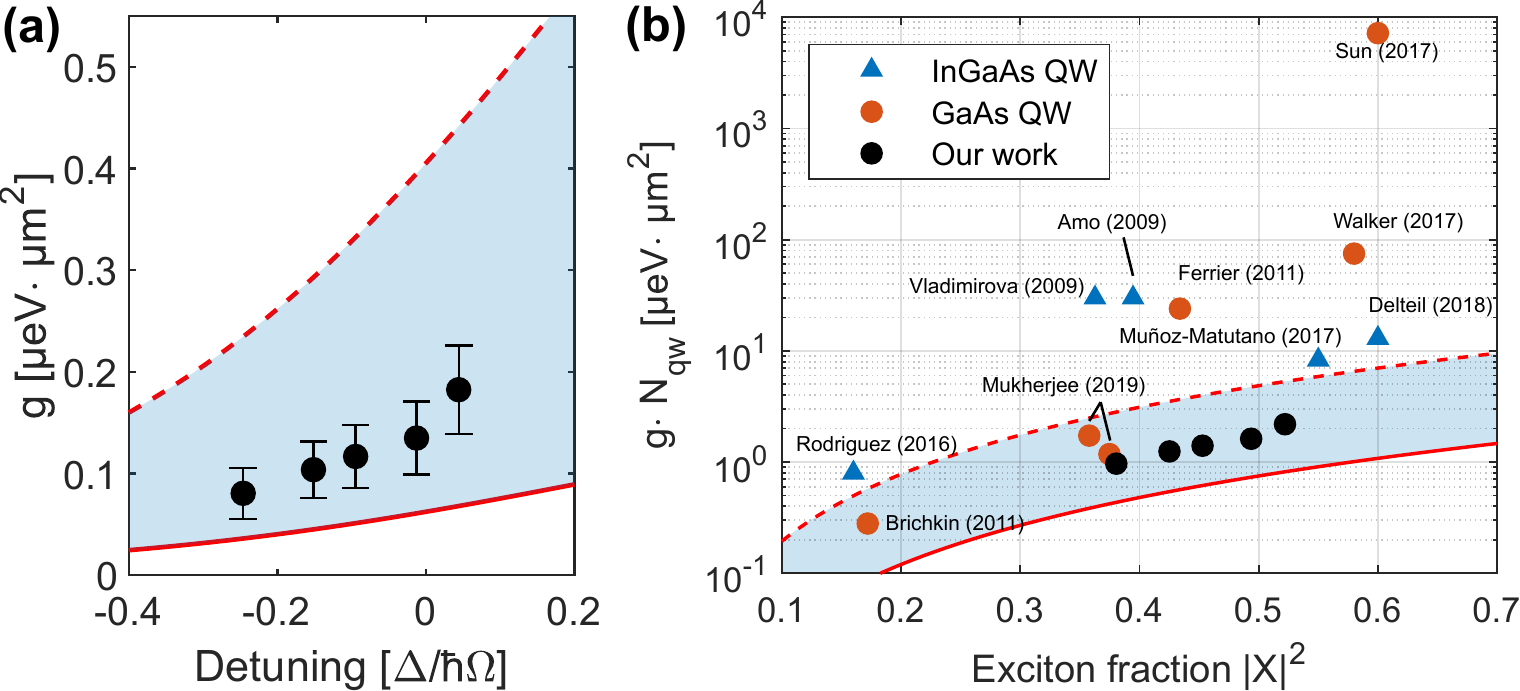}
\caption{ \textbf{Measurement of the polariton-polariton interaction strength.}
(a) Values of the polariton-polariton interaction strength extracted from the blueshift measurement shown in Fig. \ref{blueshift}(b) at different detunings $\Delta$ relative to the Rabi splitting, $\hbar\Omega$, and a constant trap diameter of $D =40$~$\mu$m. Solid curve is the theoretically predicted detuning dependence given by Eq. \ref{g}.
Upper dashed curve is the theoretical prediction of the polariton-polariton interaction in the Born approximation taking into account the 3D nature of an exciton (see text). The error bars reflect the systematic error of $~14\%$ due to density calibration (see the Supplemental Material\cite{SM}, section C) and the random error arising from density inhomogenieties of the ``flat-top" condensate (see Appendix A). The contribution of the detuning gradient to the error of this measurement is negligible. (b) Comparison of the previously reported values for the polariton-polariton interaction strength and this work (black data points).}
\label{interaction}
\end{figure*}

Figure~\ref{profiles} shows the transition from a multi-mode condensate near threshold, which has large overlap with the pump region [Fig. \ref{profiles}(a,b) and (f,g)], to a high-density ground state condensate inside the trap [Fig. \ref{profiles}(h,i)], which manifests in a clear, narrow spectral line [see Fig. \ref{profiles}(c-d)] that is observed despite the time-averaging over the duration of the pulsed experiment. This single-mode condensate has all the signatures of a spatial Thomas-Fermi distribution typical of an interacting Bose-Einstein condensate in a finite, two-dimensional circular box potential \cite{Stringari96,Vinas99,Bao16}, namely a well-defined energy (chemical potential) and a `top-hat' density distribution [Fig. \ref{profiles}(d,e)]. The shape of the potential is further corroborated by the low-energy tail in Fig. \ref{detuning}(d). The lowest energy of this tail thus represents the true bottom of the reservoir-induced potential corresponding to the polariton zero-density limit.

\section{Measurement of the polariton-polariton interaction strength}

The condensate density distribution in Fig. \ref{profiles}(d,e) and the corresponding energy blueshift can be modeled by solving the eigenvalue equation for a single-particle eigenstate of the polariton condensate in the effective potential induced by the pump-injected reservoir with the spatial density distribution $n_R(\mathbf{r})$ \cite{Wouters2007}:
\begin{equation}
-\frac{\hbar^2}{2m}\nabla^2\psi+(g|\psi|^2+g_{\rm R}n_{\rm R}+E^0_{LP})\psi =E_n\psi, 
\label{eq1}
\end{equation}
where $m$ is the effective mass of the polariton, $g$ is the polariton-polariton interaction constant, and $g_{\rm R}$ characterises the strength of interactions between the polaritons and the excitonic reservoir. Here $E_n$ is an energy eigenvalue in the effective potential $V_{\rm eff}=g|\psi|^2+g_{\rm R}n_{\rm R}$ measured relative to the minimum of the lower polariton energy in the low-density limit $E^0_{LP}$ (dashed lines in Fig.\ref{detuning}). Although, due to the cavity wedge \cite{ballistic} in our sample, $E^0_{LP}$ is a linear function of $\mathbf{r}$, this gradient can be neglected for the trap diameters used in this work, especially when the blueshift is large compared to the change of $E_{LP}^0$ across the trap.

For a trapped condensate in the ground state with chemical potential $E_0$ (measured as an off-set from $E^0_{LP}$) and a nearly uniform spatial density distribution, one can neglect the kinetic energy term in equation (\ref{eq1}), which leads to the expression: $E_0=gn_{TF}(\mathbf{r})+g_Rn_R(\mathbf{r})$. Assuming a box-like reservoir-induced potential $g_Rn_R(\mathbf{r})$ with the minimum  $n^{\rm min}_{\rm R}=0$, leads to an estimate of the interaction constant $g$ from the energy of the ground state and the local peak value of the Thomas-Fermi density distribution $g=E_0/n^{\rm peak}_{TF}$. The Thomas-Fermi approximation and the resulting estimate should apply once the polaritons condense into a single energy state and their density distribution approaches a `flat-top' profile with sharp edges, as seen in Fig. \ref{profiles}(h,i) and insets in Fig. \ref{blueshift}(a,b). The particular shape of the density distribution can be understood by recalling that the reservoir density in the pulsed excitation regime is significantly depleted by the condensate \cite{Eliezer17}, leading to a spatial domain separation, as shown in Fig. \ref{profiles}(h), and self-trapping of the condensate in a box-like effective potential. The spatial separation between the condensate and reservoir (pump) area leads to significant slowing down of gain in the polariton density with increasing pump power due to the reduction of the overlap between the condensate mode and the gain region (details on the calibration of the real-space density measurements are found in Section C of the Supplemental Material\cite{SM}). Similar condensate-induced adjustment of the effective trapping potential has been observed in the condensation of magnons \cite{magnonBEC}. 

Assuming that the only source of the energy blueshift  in the Thomas-Fermi regime (marked by points (d-f) in Fig. \ref{dispersion} (a)) is the mean-field (interaction) energy of the condensate, the polariton-polariton interaction constant can be estimated from the linear slope of the function $E_0(n^{peak}_{TF})$, as seen in the results of the numerical modeling for a fixed circular box potential presented in Fig. \ref{blueshift}(a) (see Appendix B for details). Note that at low densities, quantum confinement is the main source of the blueshift. The dependence of the ground state energy on the peak polariton density extracted from our experimental data for various diameters of the ring is shown in Fig. \ref{blueshift}(b). As the peak density of the condensate increases, its energy rapidly approaches the linear regime expected from the Thomas-Fermi limit in a fixed circular box potential (Fig. \ref{blueshift}(a)). The relative blueshift is calculated as an offset of the condensate energy from the zero-point energy in the effective potential. The latter is given by the low-energy edge of the spatially resolved spectrum, as seen in Fig. \ref{detuning}(c,d). Remarkably, in the Thomas-Fermi regime, all data points collapse onto a universal line, indicating that the relative blueshift is entirely free from the effects of quantum confinement due to the changing shape (narrowing) of the effective potential with growing pump power. However, this is not the case for a highly excitonic (positive) detuning since the trap is tight and its shape strongly deviates from a ``box'' potential, as deduced from the real-space spectra in Fig. \ref{detuning}(e,f). 

Extraction of the polariton-polariton interaction strength from the universal slope in Fig. \ref{blueshift}(b) for the detuning of $\Delta=+0.7$ meV yields the value of $g \approx 0.18 \pm 0.018$ $\mu$eV$\cdot \mu$m$^2$. Furthermore, we have extracted different values of the slope $\Delta E_0/\Delta n^{peak}_{TF}$ for varying detuning by pumping with a fixed ring size ($D=40$ $\mu$m) at different positions on the sample, as shown in Section D of the Supplemental Material\cite{SM}. The values of the polariton-polariton interaction strength, $g$, measured in the Thomas-Fermi regime are shown in Fig. \ref{interaction}(a). As expected, the strength of interaction grows with the larger proportion of exciton in the polariton quasiparticle. Note that for this measurement, we are limited to the near-zero detuning range where we can create a flat-top condensate in a circular box potential, hence the limited set of the data points in Fig. \ref{interaction}(a). At more negative (photonic) detunings, the condensate is fragmented, making it impossible to define a single chemical potential. On the other hand, at more positive (excitonic) detunings, there is a significant presence of reservoir particles inside the ring [see Fig. \ref{detuning}(e,f)]. This results in a larger blueshift, which is already evident in the largest positive detuning data point of Fig. \ref{interaction}(a). Further buildup of the polariton density inside the optically-induced trap leads to a departure from the Thomas-Fermi regime, as seen from the dramatic rise of the blueshift in Fig. \ref{blueshift}(b) for the highest densities. This regime is reached for lower peak densities of polaritons in smaller-area traps and for larger peak densities in larger-area traps, thus indicating a strong influence of quantum confinement. Even though the densities per QW in this regime are only an order of magnitude lower than the Mott density in the $7$-nm QWs, strong coupling is retained, which indicates that the observed behaviour could be attributed to a BEC-BCS crossover \cite{Keeling,Byrnes10,BCS_njp,triplet}. However, detailed analysis of this regime is beyond the scope of the present study.

It is possible to compare our results with the theoretical prediction derived in the Born approximation \cite{ciuti98,tassone,Rochat,Glazov}. As pointed out in  \cite{Forchel11,CiutiREV13}, and summarised in section E of the Supplemental Material\cite{SM}, the effective contact polariton-polariton interaction constant entering Eq. \ref{eq1} has the following main contribution:  
\begin{equation}
g=|X|^4 g_{\rm X}/(2N_{\rm QW}),
\label{g}
\end{equation}
where $g_{\rm X}$ is the exciton-exciton interaction constant, and the factor of $2$ accounts for the dominant role of triplet interactions. The Hopfield coefficient  $|X|^2 = (1/2)\left(1+\Delta/\sqrt{\Delta^2 + \hbar^2\Omega^2}\right)$, where $\hbar \Omega$ is the Rabi splitting, determines the exciton fraction in a polariton, and $N_{\rm QW}$ is the number of quantum wells in the sample. Using the theoretical estimate for the exciton-exciton interaction constant \cite{ciuti98,tassone,Glazov} $g_{\rm X}=6E_Ba^2_B$, with the realistic values for the exciton binding energy $E_B\approx10$ meV and the Bohr radius $a_B\approx 10$ nm, we obtain $g_{\rm X}\approx 6 ~\mu$eV$\cdot \mu$m$^2$. Using this value for $g_{\rm X}$ in Eq. \ref{g}, we obtain the solid curve in Fig. \ref{interaction}(a), which sits below our data points. However, as noted in Section E of the Supplemental Material\cite{SM}, the conventional expression for $g_{\rm X}$ is derived by using a purely 2D exciton wavefunction. In practice, the 3D nature of an exciton may become important since the Bohr radius is comparable to the width of a quantum well. Taking into account the 3D nature of excitons (see Supplemental Material\cite{SM}, Section E), yields an upper bound on the theoretical value of polariton-polariton interaction strength shown by the dashed curve in Fig. \ref{interaction}(a). 

An alternative source of discrepancy between the theory and experiment suggested by several studies is the effect of saturation of the exciton oscillator strength at larger densities \cite{saturation,tassone,Rochat,Glazov}, which effectively manifests itself as an additional  contribution to the interaction strength (\ref{g}): $\Delta g_{\rm sat}=|X|^2XCg_{\rm sat}/(2N_{\rm QW})$, where $|C|^2=1-|X|^2$ is the photon fraction of the polariton. The saturation correction is typically assumed to be small and omitted from consideration \cite{CiutiREV13}, however for our experiment with a large Rabi splitting and large densities, the contribution of this term can be significant. With the saturation correction as estimated in \cite{Rochat,Forchel11}, the theoretical value for $g$, would, in principle, show an excellent agreement with our experimental data. However, this correction to the interaction strength has been estimated within a perturbative framework that assumes that the exciton wave function is unaffected by the light-matter coupling \cite{saturation,tassone,Rochat,Glazov}. This assumption is not justified in our system since the Rabi splitting is comparable to the exciton binding energy. Indeed, light-induced changes to the exciton radius have already been predicted \cite{khurgin2001} and measured \cite{hoefling2017}. Consequently, we find that the saturation term predicted in \cite{Rochat,Forchel11} overestimates the effect of the exciton oscillator strength saturation \cite{levinsen2019}, and other perturbative corrections to the interaction strength may play a greater role in this regime. 

To compare the measurement presented in Fig. \ref{interaction}(a) with previous results, in Fig. \ref{interaction}(b) we plot the previously reported values of the polariton-polariton interaction strength scaled by the number of quantum wells as a function of the exciton fraction, the latter determined from the detuning and the Rabi splitting. Our results are shown in Fig. \ref{interaction}(b) (black circles) in comparison with the data at various fixed detuning (exciton fraction) values reported for the microcavities with GaAs \cite{Forchel11,Bloch11,Walker17,Nelson17,Mukherjee19} and InGaAs \cite{Amo09,Vladimirova09,Rodriguez16,Volz17,Deitel18} quantum wells. One can see that the range of the values obtained by various methods and reported in the literature spans four orders of magnitude, and most of them exceed the conventional theoretical estimate (solid line) by at least an order of magnitude. The simple comparison of different available data in this figure with the range of theoretical values does not capture the dependence of the interaction strength on the Rabi splitting and the slightly different exciton properties in the two materials, however corrections to the theory curves introduced by these effects are not significant on the logarithmic scale of the plot. Arguably, the measurements that derive the interaction strength from the density-dependent blueshift involve the least number of fitting parameters, provided the density calibration is accurate. However, as discussed in \cite{Walker17}, these measurements are performed on trapped condensates \cite{Forchel11,Bloch11,Walker17} and typically produce values larger than those predicted theoretically due to the quantum confinement effect and the fact that complete separation between the polariton and reservoir density cannot be achieved in optically-induced traps created in a {\em cw} regime, making it difficult to eliminate polariton-reservoir interaction. The extreme case of these effects is represented by the value extracted from the measurements performed below the condensation threshold in optically-induced traps in the samples similar to that used here \cite{Nelson17} [the uppermost point in Fig. \ref{interaction}(b)]. We argue that this latter measurement cannot be used to reliably determine the polariton-polariton interaction strength since the energy blueshifts below threshold are solely due to strong polariton-reservoir interaction \cite{Pieczarka18}. Furthermore, recent values of the large interaction strength deduced from the correlation measurements in the resonant excitation regime (see, e.g., Ref. \cite{Deitel18}) are also likely to be affected by the presence of a reservoir \cite{Stepanov18}. By contrast, our measurement in the Thomas-Fermi regime fully eliminates the influence of the incoherent excitonic reservoir.  

\section{Conclusions} 
To summarise, we have observed the Thomas-Fermi wavefunction of a polariton condensate in a box potential corresponding to the true ground state of the trapped exciton-polariton system. This, to our knowledge, is the first time the single-mode Thomas-Fermi regime has been demonstrated in any non-equilibrium quantum coherent system. The Thomas-Fermi regime was reached in our experiment by employing a spatial hole burning effect in a single-shot realisation of a long-lifetime condensate in a circular optically-induced potential, whereby complete spatial separation was achieved between the condensate and the thermal reservoir. Although a degree of separation between a condensate and an optical excitation region has been demonstrated within optical traps in the regime of {\em cw} excitation (see, e.g., \cite{Bloch11,Snoke}), the full spatial depletion of the reservoir enabled by the pulsed excitation in our experiments is critical for removing the spatial overlap between the polaritons and the reservoir. On the other hand, optical trapping is essential for reaching the high-density Thomas-Fermi regime since, without a trap, condensates that are sufficiently separated from the excitation region have low density (see, e.g., \cite{Sanvitto17}). 

By driving condensation into this interaction-dominated regime, we were able to apply  a `textbook' local density approximation, and extract the values of the polariton-polariton interaction strength in very good agreement with theoretical estimates \cite{ciuti98,tassone,Rochat,Glazov}. We argue that our work represents a direct and conclusive measurement of the polariton-polariton interaction strength, and therefore serves to reduce the four orders of magnitude uncertainty in this key parameter of exciton-polariton physics. Our work therefore shows that an exciton-polariton condensate should be treated as a weakly interacting quantum gas, and points to the polariton-reservoir interaction as a key factor contributing to previously reported larger-than-predicted values of polariton-polariton interaction strength.

Furthermore, the polariton condensate in the Thomas-Fermi regime has a nearly-uniform `flat-top' density profile (neglecting the boundary regions), which allows studies of the condensate without strong density gradients, in analogy with ultracold atomic gases in a box trap \cite{BECbox,Dalibard14,Hadzibabic14,qdepletion17}. Such spatially homogeneous condensates are extremely important for the investigations of fundamental properties of Bose-Einstein condensation, since they enable direct comparison with theoretical and analytical results, usually derived for homogeneous cases. Thus, the Thomas-Fermi condensate observed here provides a platform for novel studies of a 2D non-equilibrium quantum system, such as the relationship between coherence and dynamically or thermally excited topological defects \cite{Dalibard14}, critical dynamics of symmetry breaking upon transition to condensation \cite{Navon15}, elementary excitations \cite{Yamamoto,Stepanov18} and quantum depletion \cite{qdepletion,qdepletion17}. 

\section*{Appendix A: Experimental details and the sample}

The high-Q microcavity sample used in this work consists of $12$ (three groups of four) $7$-nm GaAs quantum wells embedded in a $3\lambda/2$ cavity composed of $32$ (top) and $40$ pairs (bottom) of Al$_{0.2}$Ga$_{0.8}$As/AlAs layers of distributed Bragg reflectors; the exact sample used in \cite{Eliezer17} and similar to \cite{SnokePRX13,Snoke,Nelson17}. The Rabi splitting is $\hbar\Omega=15.85$ meV, the exciton energy is $E_X=1606.2$ meV, and the cavity photon mass is $\approx 3.4\times10^{-5}$ of the free electron mass. All experiments are performed below $10$ K using a continuous flow microscopy cryostat.

In the experimental setup, we use a mode-locked Ti:Sapphire laser (Chameleon Ultra II, $80$ MHz of $140$-fs pulses) modulated by an acoustic-optical modulator ($1-10$ kHz at $10$-$\mu$s high-time) to off-resonantly pump the system with linearly polarized light at wavelengths below $730$ nm. A $50\times$ objective (${\rm NA}=0.5$) focuses the laser onto the sample and collects the cavity photoluminescence. Two methods are used to shape the pump beam. Small rings are created with an amplitude mask made of a milled metal shim reimaged onto the sample using a tube lens and the objective. For larger rings, we use a $1^\circ$ axicon lens placed between two confocal plano-convex lenses before the objective. A free-space microscope reimages the near-field (real space) and far-field (momentum space) onto a camera, while a monochromator enables energy-resolved measurements. 

The experiments presented here are performed in the pulsed regime, i.e. by exciting the polaritons with a $140$-fs pulse which arrives with a $12.5$-ns repetition period. This is significantly longer than the decay time of both polaritons ($\sim200$ ps) and the reservoir ($\sim1$ ns) in this sample. The single-shot imaging presented in Section B of the Supplemental Material\cite{SM} is enabled by a homebuilt high-contrast ratio (1:10,000) pulse-picker \cite{Eliezer17} that picks single pulses from the mode-locked laser synced to an electron-multiplying CCD (Photometrics Cascade 512b) which is exposed for at least $10$ $\mu$s before and after the pulse. A single-shot real-space image is time-integrated over the whole duration of the photoluminescence emission. 

The peak densities of the condensate in the Thomas-Fermi regime are extracted from the density profiles corresponding to the maxima of the corresponding spectral lines. Due to inhomogeneity of the sample and imperfections of the imaging setup, the condensate profiles are not perfectly flat, as shown in the inset of Fig. \ref{blueshift}(b). The spatial distribution of the polariton density in our sample is also affected by the cavity wedge \cite{ballistic}, which shifts the peak density of the ground state towards one side of the trap. This adds an uncertainty to the actual peak density, therefore we define a lower bound for the measured peak density as the average density of the regions with $n/n_{max} > 1-1/e$ and an upper bound for regions with $n/n_{max} > 0.95$.

\section*{Appendix B: Modeling} 
To model the formation and decay of the condensate produced by a single laser pulse, we employ the open-dissipative Gross-Pitaevskii model \cite{Wouters2007} with a phenomenological energy relaxation responsible for the effective reduction of the chemical potential of the condensate  \cite{Wouters_relaxation}, and an additional stochastic term accounting for fluctuations \cite{Wouters2009}, as outlined in \cite{Eliezer17}:
\begin{widetext}
\begin{eqnarray}
i\hbar\frac{\partial\psi(\mathbf{r})}{\partial t}&=&\left[ (i\beta-1)\frac{\hbar^2}{2m}\nabla^2+g|\psi|^2+g_{\rm R}n_{\rm R}+i\frac{\hbar}{2}\left(Rn_{\rm R}-\gamma\right)\right]\psi(\mathbf{r}) + i\hbar\frac{dW}{d t}, \nonumber
\\
\frac{\partial n_{\rm R}(\mathbf{r})}{\partial t}&=&-(\gamma_{\rm R}+R|\psi(\mathbf{r})|^2)n_{\rm R}(\mathbf{r})+P(\mathbf{r}).
\nonumber
\end{eqnarray}
\end{widetext}
Here, $R$ defines the stimulated scattering rate, $\gamma$ and $\gamma_R$ are the decay rates of condensed polaritons and the reservoir, respectively. The rate of injection of the reservoir particles, $P(\mathbf{r})$, is proportional to the pump power, and its profile is defined by the pump. The term $\propto dW/dt$ introduces a stochastic noise in the form of a Gaussian random variable with the white noise correlations: 
\begin{equation}
\langle dW^{*}_idW_j\rangle= \frac{\gamma+R n_{\rm R}(\textbf{r}_i)}{2(\delta x \delta y)^2} \delta_{i,j}dt, \qquad \langle dW_idW_j\rangle=0, \nonumber
\end{equation}
where $i$, $j$ are discretisation indices: $\mathbf{r}_i=(\delta x$,$\delta y)_i$. The parameters defining the time scales for radiative decay and thermalisation processes, $\gamma$, $\gamma_R$, $R$, and $\beta$, are varied consistently with the characteristic values for long-life polaritons at various exciton-photon detunings. A single-shot realisation of the condensation process corresponds to a single realisation of the stochastic process modelled by these equations, and real-space images of the condensate distribution obtained using this model agree with the experiment remarkably well (see Fig. \ref{profiles} and Supplemental Material \cite{SM}, Section B).

To demonstrate the typical features of a transition from the regime dominated by the quantum confinement at low density to the interaction-dominated Thomas-Fermi regime, we use the same model equations to determine a steady-state condensate wavefunction in a circular potential well $V({\bf r})$ with a diameter $D=20$ $\mu$m and a depth of $5$ meV. The shape of the well is fixed and independent on the density of the trapped polaritons. To inject polaritons into the well, a $5$ $\mu$m pump is focused at the centre of the trap. The blueshift of the ground state as a function of peak density is shown in Fig. \ref{blueshift}(a). The simulation clearly shows the zero point energy offset due to the quantum confinement and a slow increase in blueshift at low density. With increasing density, the blueshift eventually follows the Thomas-Fermi limit $\Delta E=g n^{\rm peak}$, as observed in the experiment, albeit at a higher density.

\end{document}